\documentclass[a4paper]{article}

\usepackage{authblk}
\usepackage{atlasphysics}
\usepackage{placeins}
\usepackage{xcolor}
\usepackage{xspace}
\usepackage{verbatim}
\usepackage{subfigure}
\usepackage{graphicx}
\usepackage{comment}
\usepackage{url}
\usepackage[margin=1.0in]{geometry}
\usepackage[hyperfootnotes=true]{hyperref}
\graphicspath{{}{figures//}{plots//}}

\def\ttH{\ensuremath{t\bar{t}H}\xspace}
\def\Hmumu{\ensuremath{H\rightarrow\mu^+\mu^-}\xspace}
\def\ttZ{\ensuremath{t\bar{t}Z}\xspace}
\def\ttW{\ensuremath{t\bar{t}W}\xspace}
\def\ttWW{\ensuremath{t\bar{t}WW}\xspace}
\def\tt{\ensuremath{t\bar{t}}\xspace}
\def\pt{\ensuremath{p_{\mathrm{T}}}\xspace} % Subscript roman not italic (EE)
\def\pT{p_{\rm{T}}} % Subscript roman not italic (EE)
\def\HT{\ensuremath{H_{\mathrm{T}}}\xspace}
\def\jfakemu{\ensuremath{f_{j\rightarrow\mu}}}

\title{Study of \ttH ($\Hmumu$) in the\\ three lepton channel at $\sqrt{s} = 14$ TeV;\\
A Snowmass white paper} % Article title
\author[a]{Jahred Adelman, Andrey Loginov, Paul Tipton, Jared Vasquez}
\affil[a]{Yale University}
\date{}

\usepackage{lineno}
\begin{document}

\maketitle
\begin{abstract}
The $\Hmumu$ signature provides excellent mass resolution for Higgs
bosons, and is therefore an important Higgs boson decay channel despite
the small dimuon branching ratio. 
We present an optimization of selection criteria in a search for trilepton 
\ttH ($\Hmumu$) events, in which the top quark pair decays
semi-leptonically, at a simulated High Luminosity LHC (HL-LHC) running
at 14 TeV. The study is performed with 3000 fb$^{-1}$ of simulated
data with an average pileup of $<\mu> = 140$. In this ultimate HL-LHC
data set, we find that \ttH ($\Hmumu$) will be a very difficult
signature to observe due to the very small expected signal.
\end{abstract}

%
%%%%%%%%%%%%%%%%%%%%%%%%%%%%%%%%%%%%%%%%%%%%%%%%%%%%%%%%%%%%%%%%%%%%%%%%%%%%%%%
% Introduction
%%%%%%%%%%%%%%%%%%%%%%%%%%%%%%%%%%%%%%%%%%%%%%%%%%%%%%%%%%%%%%%%%%%%%%%%%%%%%%%
%

\section{Introduction}

The particle with mass around 125 GeV recently observed by both the
ATLAS~\cite{higgsATLAS} and CMS~\cite{higgsCMS} collaborations is
compatible~\cite{Aad:2013wqa,CMS:yva,Aad:2013xqa} with production of
the Standard Model (SM) Higgs boson. The particle nearly completes the
SM, and explains both how elementary particles obtain mass as well as
electroweak symmetry
breaking~\cite{higgsthry1,higgsthry2,higgsthry3,higgsthry4,higgsthry5,higgsthry6}.
Nevertheless, one of the most important questions in particle physics
is whether the discovered particle is the Higgs boson predicted by the
SM or a similar impostor.

\begin{figure}[!htbp]
\centering
\includegraphics[scale=0.4]{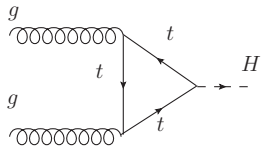}
\caption{Representative Feynman diagram of Higgs boson production via gluon fusion. }
\label{fig:ggH_feyn}
\end{figure}

\begin{figure}[!htbp]
\centering
\includegraphics[scale=0.3]{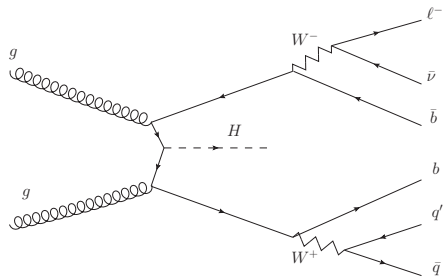}
\caption{Feynman diagram of $\ttH$ production.}
\label{fig:ttH_feyn}
\end{figure}

The dominant mode for Higgs boson production in proton-proton ($pp$)
collisions at the LHC is $gg \rightarrow H$, known as gluon fusion.
In gluon fusion, the Higgs boson is produced through a quark
(primarily top-quark) loop, as shown in
Fig.~\ref{fig:ggH_feyn}. However, only events with the Higgs boson
produced with a top-quark pair ($t\bar{t}$) allow the direct
observation and study of the \ttH vertex, as shown in
Fig.~\ref{fig:ttH_feyn}.

At the observed mass near 125 GeV the Higgs boson will decay most
prominently into a pair of bottom quarks as shown in
Table~\ref{h_decays.table}. The Higgs boson may also decay into a muon
pair ($\Hmumu$), although with a much smaller branching ratio of
$2.20\times10^{-4}$. The decay benefits from having a very clear
$\Hmumu$ mass resolution, similar to the $H \rightarrow \gamma\gamma$
decay, whose excellent mass resolution made it vital to the Higgs
boson discovery despite its small branching ratio. To reduce the
background levels, the analysis is performed in trilepton \ttH events,
where one top-quark decays leptonically to either an electron or muon,
which we shall call the electron and muon channel, respectively. An
event with the aforementioned production and decay mode in the
trilepton channel would be seen in the detector as 4 jets (including
the 2 $b$-jets), two oppositely charged muons, an additional lepton
and a neutrino from the $W$-boson decay, inferred by missing
transverse momentum (\MET{}).

\begin{table}
\begin{tabular} {c|c|c|c|c|c|c|c|c|c} 
%\hline
Channels & 
$b\bbar$        & 
$\tau^+\tau^-$     &  
$\mu^+\mu^-$       & 
$c\bar{c}$     &  
$gg$           & 
$\gamma\gamma$ &  
$Z\gamma$      &  
$WW$           & 
$ZZ$           \\ 
\hline
BR &
0.58            &
0.063            &
2.20 x 10$^{-4}$ &
0.0291 &
0.086 &
2.28 x 10$^{-3}$ &
1.54 x 10$^{-3}$ &
0.215 &
0.026 \\
%\hline
\end{tabular}
\caption{Branching ratios for the Higgs boson decay modes for $m_H = 125$~GeV.}
\label{h_decays.table}
\end{table}

%\begin{figure}[!htbp]
%\centering
%\includegraphics[scale=0.55]{Higgs_proXS_14TeV}
%\caption{Cross sections for different Higgs boson production models as a function of Higgs boson mass, taken from Ref.~\cite{LHCHiggsWG}. }
%\label{fig:prodM}
%\end{figure}

%\begin{figure}[!htbp]	
%\centering
%\includegraphics[scale=0.55]{Higgs_BR_LM}
%\caption{Higgs boson branching ratios as a function of Higgs boson mass, taken from Ref.~\cite{LHCHiggsWG}. }
%\label{fig:decayM}
%\end{figure}

%
%%%%%%%%%%%%%%%%%%%%%%%%%%%%%%%%%%%%%%%%%%%%%%%%%%%%%%%%%%%%%%%%%%%%%%%%%%%%%%%
% Modeling and Object Selection
%%%%%%%%%%%%%%%%%%%%%%%%%%%%%%%%%%%%%%%%%%%%%%%%%%%%%%%%%%%%%%%%%%%%%%%%%%%%%%%
%

\section{Modeling and Object Selection}

This study was done in the context of Snowmass with official Snowmass
samples using the Delphes fast simulation framework~\cite{Delphes} of
a generic High Luminosity LHC (HL-LHC) detector with particle flow to
mitigate the pileup. 
%Backgrounds were
%generated in MadGraph and simulated in the Delphes parametrized
%FastSimulation framework 
The study focuses on a HL-LHC running at 14 TeV, and a combined
integrated luminosity of 3000 fb$^{-1}$ with an average pileup of
$<\mu> = 140$.

\begin{comment}
"Delphes is a C++ framework, performing a fast multipurpose detector
  response simulation. The simulation includes a tracking system,
  embedded into a magnetic field, calorimeters and a muon system. The
  framework is interfaced to standard file formats (e.g. Les Houches
  Event File or HepMC) and outputs observables such as isolated
  leptons, missing transverse energy and collection of jets which can
  be used for dedicated analyses. The simulation of the detector
  response takes into account the effect of magnetic field, the
  granularity of the calorimeters and subdetector
  resolutions. Visualization of the final state particles is also
  built-in using the corresponding ROOT library."
  (http://arxiv.org/pdf/1307.6346v1.pdf)
\end{comment}

In the Delphes framework identified leptons (electrons and muons) are
required to be isolated~\cite{Avetisyan:2013onh}. To be accepted as
`good', electrons (muons) are required to have $|\eta| <
2.5~(2.4)$. Jets are required to have $|\eta| < 2.7$. After comparing
the \pt spectra of electrons, muons and jets, good leptons are
required to have a minimum \pt of 25 GeV and good jets are required to
have a minimum \pt of 30 GeV~\cite{Avetisyan:2013onh}. These cuts were
chosen to maintain signal efficiency while reducing the background as
much as possible. Further optimization of the $\ttH$ event selection
have been performed, as described in Section~\ref{s:selection}.

%
%%%%%%%%%%%%%%%%%%%%%%%%%%%%%%%%%%%%%%%%%%%%%%%%%%%%%%%%%%%%%%%%%%%%%%%%%%%%%%%
% Background Samples
%%%%%%%%%%%%%%%%%%%%%%%%%%%%%%%%%%%%%%%%%%%%%%%%%%%%%%%%%%%%%%%%%%%%%%%%%%%%%%%
%
\subsection{Signal Sample}
A signal sample of $\ttH, \Hmumu$ was generated using
MadGraph5~v2~beta~\cite{Maltoni:2002qb} with showering provided by
Pythia6~\cite{Sjostrand:2006za}. The sample was processed through the
standard Delphes simulation as all background processes.

\subsection{Background Samples}

Backgrounds to the trilepton $\ttH, \Hmumu$ signature are processes
with prompt leptons and $b$-jets, as well as processes with at least
one misidentified (`fake') object. For instance, a light-flavor jet
can be misidentified as a heavy-flavor jet. A lepton that comes from a
non-prompt (non-$W/Z$) source or from a misidentified hadron, can also
pass the selection criteria. Background processes containing real and
misidentified leptons and $b$-jets were considered in this analysis.

The background samples and cross sections from the official Snowmass
Energy Frontier generation~\cite{Avetisyan:2013onh,Anderson:2013kxz}
were used for this analysis. Simulation samples of $V$ + jets ($V$ =
$W$, $Z^0/\gamma^*$) production, $VV$ and $VVV$ production, single top
and top-quark pair production were generated in $\HT$ bins (where
$\HT$ is defined as the scalar sum of jet and lepton transverse
momenta), as described in Ref.~\cite{Avetisyan:2013onh}.

One of the dominant backgrounds is top-quark pair production in
association with a boson, i.e. \ttZ, \ttW, and \ttWW. These samples
were generated using MadGraph5~v2~beta~\cite{Maltoni:2002qb} using
Pythia6~\cite{Sjostrand:2006za} showering. Cross sections for the
signal and $\ttbar$+X background samples are summarized in
Table~\ref{t:topb}.

% Backgrounds are grouped into categories: (a) $V$ + jets ($V$ = $W$,
% $Z^0/\gamma^*$) (Table~\ref{t:bosonplusjets}); (b) $VV$ and $VVV$
% samples (Table~\ref{t:diboson}); (c) single top and top-quark pair
% production samples (Table~\ref{t:top}); and (d) top-quark pair
% production in association with a gauge boson
% (Table~\ref{t:topb}). These samples were generated in $\HT$ bins
% (where $\HT$ is defined as the scalar sum of jet and lepton
% transverse momenta).

\begin{table}[!htbp]
\centering
\begin{tabular}{l | l} 
%\hline\hline
Process & Cross Section, pb \\
\hline
%\hline
\ttH   &  0.6113~\cite{Dittmaier:1318996} \\
\hline
\ttWW  &  0.0104~\cite{Maltoni:2002qb} \\
\ttW   &  0.7062~\cite{Maltoni:2002qb} \\
\ttZ   &  0.0741~\cite{Maltoni:2002qb} \\
\ttbar &  953.6~\cite{Czakon:2013goa}  \\
\end{tabular}
\caption{Cross sections for the inclusive \ttH production and for the
  major backgrounds. The \ttH cross section is the value before the
  $\Hmumu$ branching ratio is applied.}
\label{t:topb}
\end{table}

%
%%%%%%%%%%%%%%%%%%%%%%%%%%%%%%%%%%%%%%%%%%%%%%%%%%%%%%%%%%%%%%%%%%%%%%%%%%%%%%%
% Muon Fakes
%%%%%%%%%%%%%%%%%%%%%%%%%%%%%%%%%%%%%%%%%%%%%%%%%%%%%%%%%%%%%%%%%%%%%%%%%%%%%%%
%
\subsection{Muon Fakes}

Inclusive \tt production (with at least one fake lepton) is one of the
dominant backgrounds for this analysis. To mitigate large statistical
uncertainties on the prediction of this background in the muon
channel, a per-jet $j\rightarrow\mu$ fake rate probability is applied
to every \tt dimuon event.

%al: do we need these details? 
% For each event, this is applied 2000 times and then averaged.

The fake rate $\jfakemu$ was derived by taking the number of events
with a fake muon and dividing that by the number of jets that could
fake a muon. To estimate the number of fake leptons, we assume that
every event in the same-sign (SS) dimuon channel of \tt has a fake
lepton. We further assume that a fake muon is equally likely to have
positive or negative charge, so we multiply our count of SS dimuon
events by a factor of two to account for the number of fake muons in
the opposite sign (OS) dimuon channel. Finally, we divide this number
by the total number of jets with minimum \pt of 25~GeV in the
semi-leptonic muon channel of the \tt sample to find the fake rate as
follows:

\begin{equation}
\jfakemu = \frac{ 2 \times N_{SS} }{ N_{jets~in~\mu+jets} } = 6.5 \times 10^{-5}
\end{equation}

\begin{figure}[!htbp]	
\centering
\includegraphics[scale=0.55]{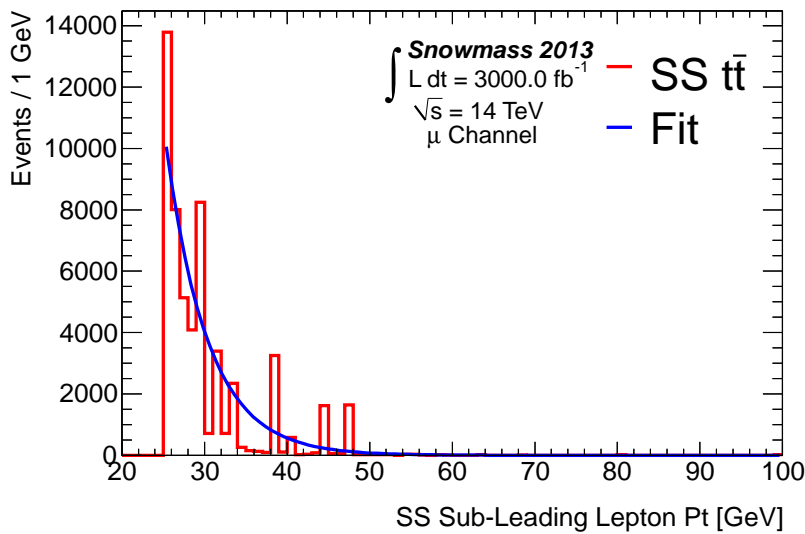}
\caption{Distribution and fit of the fake muon \pt derived from same-sign dimuon channel of \tt. }
\label{fig:fakept}
\end{figure}

The \pt distribution for fake muons is also taken from the SS dimuon
channel of \tt. The sub-leading muon is assumed to be the fake muon.
The sub-leading muon $\pt$ distribution is fit to an exponential curve
as shown in Fig.~\ref{fig:fakept}. The fit is used at random as a
probability density for the fake muon \pt. The direction ($\eta$ and
$\phi$) are taken from the jet which will be the fake muon.
%al: too many details:
% Jets that are counted as fake muons are not counted as jets.

%
%%%%%%%%%%%%%%%%%%%%%%%%%%%%%%%%%%%%%%%%%%%%%%%%%%%%%%%%%%%%%%%%%%%%%%%%%%%%%%%
% Event selection
%%%%%%%%%%%%%%%%%%%%%%%%%%%%%%%%%%%%%%%%%%%%%%%%%%%%%%%%%%%%%%%%%%%%%%%%%%%%%%%
%
\section{Event selection}
\label{s:selection}

In this study, we focus on the \ttH signature where the Higgs boson
decays to dimuons and the top-quark pair decays semi-leptonically to
either an electron or muon, which we shall call the electron and muon
channel, respectively. Each channel was optimized independently,
however the two optimizations gave identical cuts.

\begin{figure}[!tbhb]	
%\centering
\mbox{
\subfigure[\small Leading muon $\pT$, muon channel]{\includegraphics[scale=0.38]{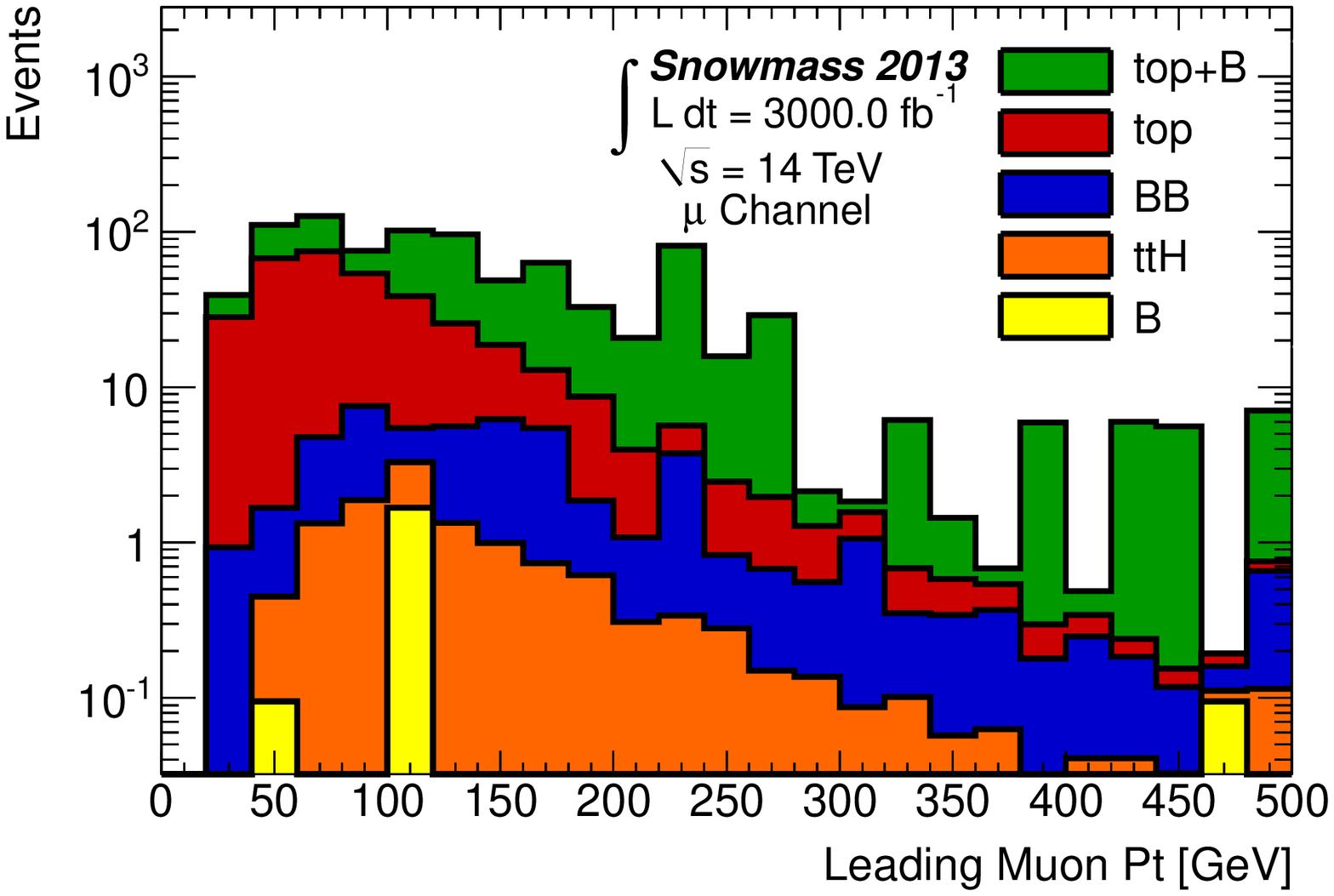}} %both are taken from combined_m4.root
\subfigure[\small Leading muon $\pT$, electron channel]{\includegraphics[scale=0.38]{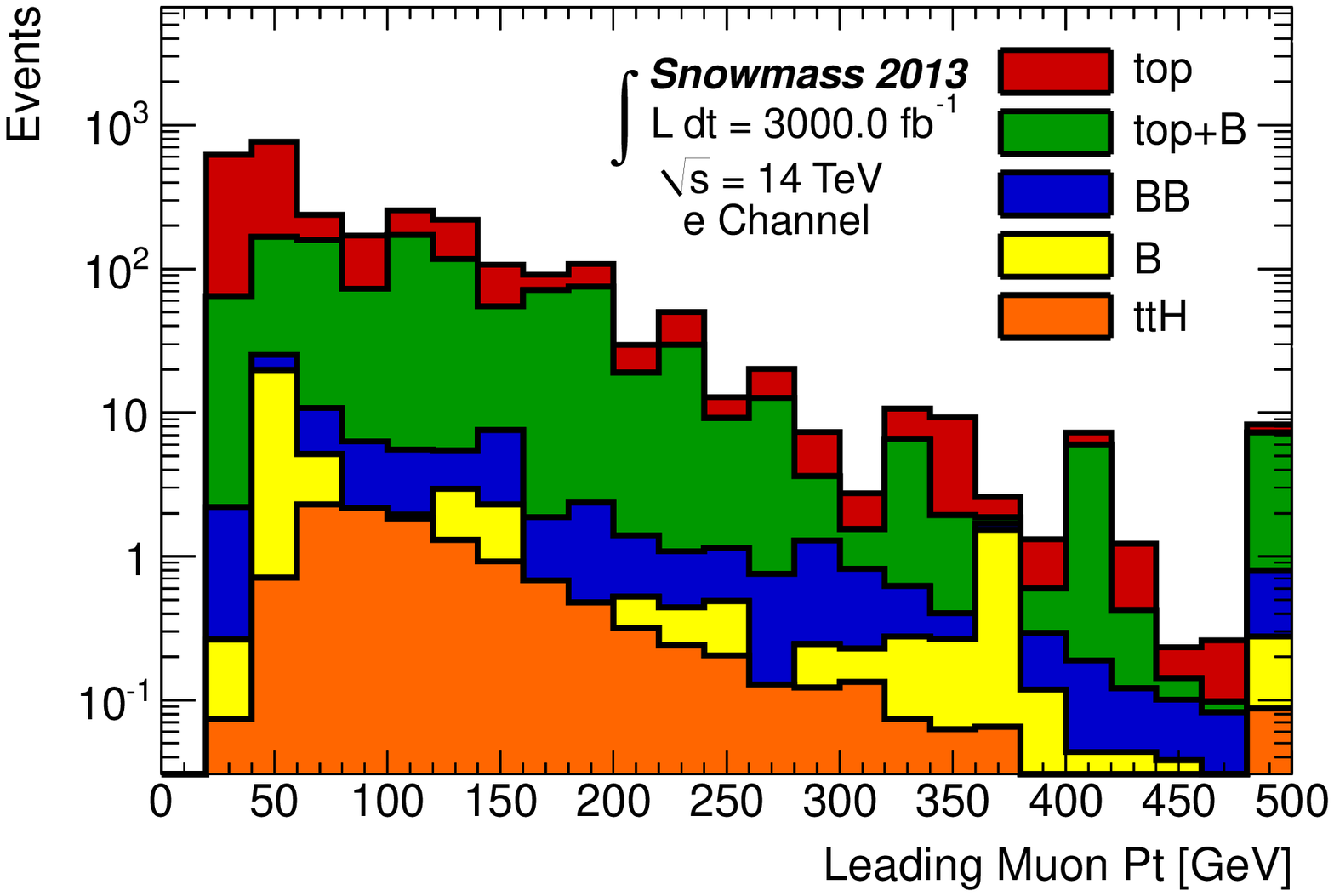}}}
%\caption{Leading muon \pt distribution for the muon (left) and electron (right) channels.}
%\label{fig:l1pt_precut}
%\end{figure}
%\begin{figure}[!htbp]	
%\centering
\mbox{
\subfigure[\small \HT distribution, muon channel]{\includegraphics[scale=0.38]{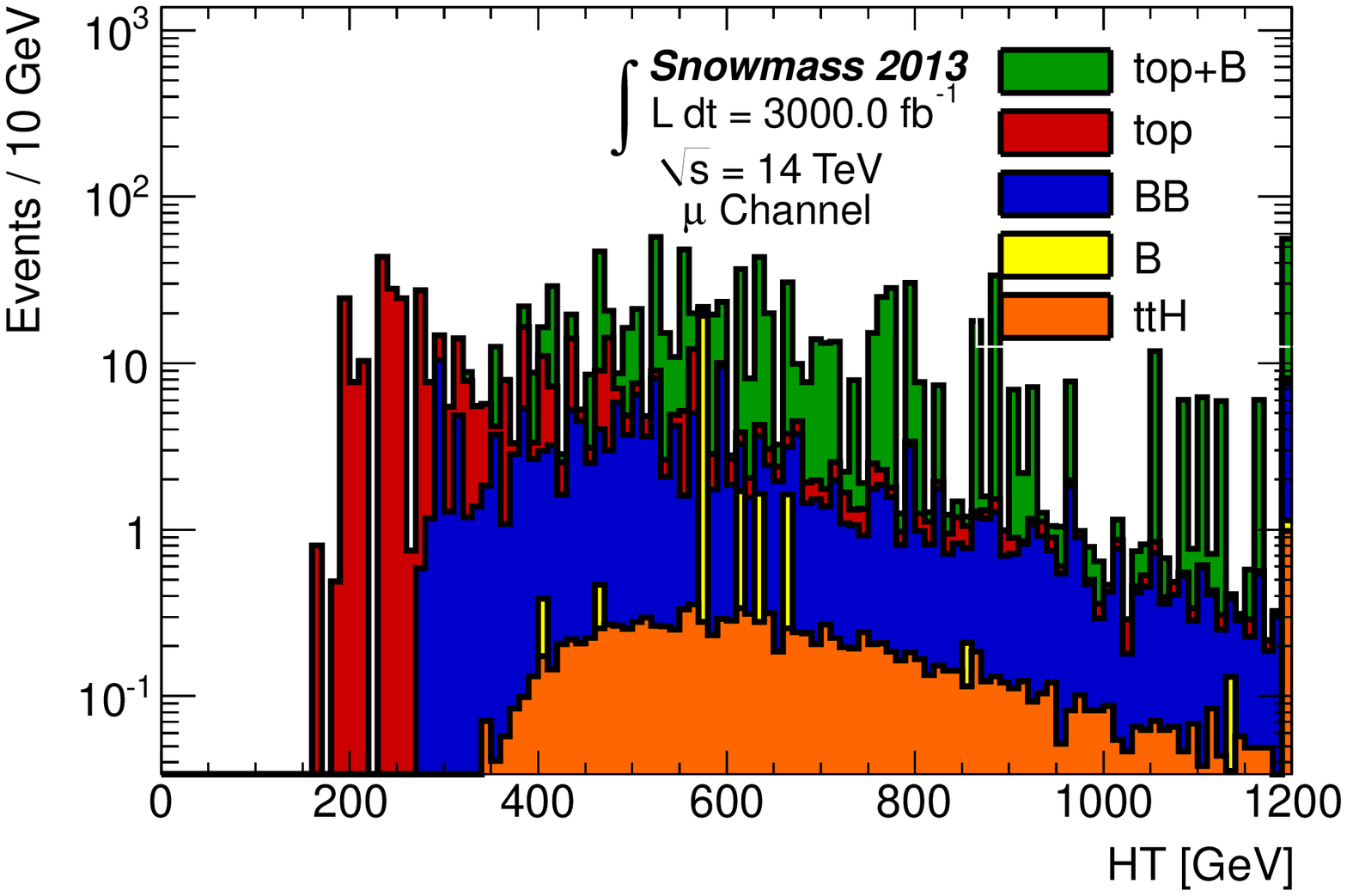}} %both are taken from combined_m4.root
\subfigure[\small \HT distribution, electron channel]{\includegraphics[scale=0.38]{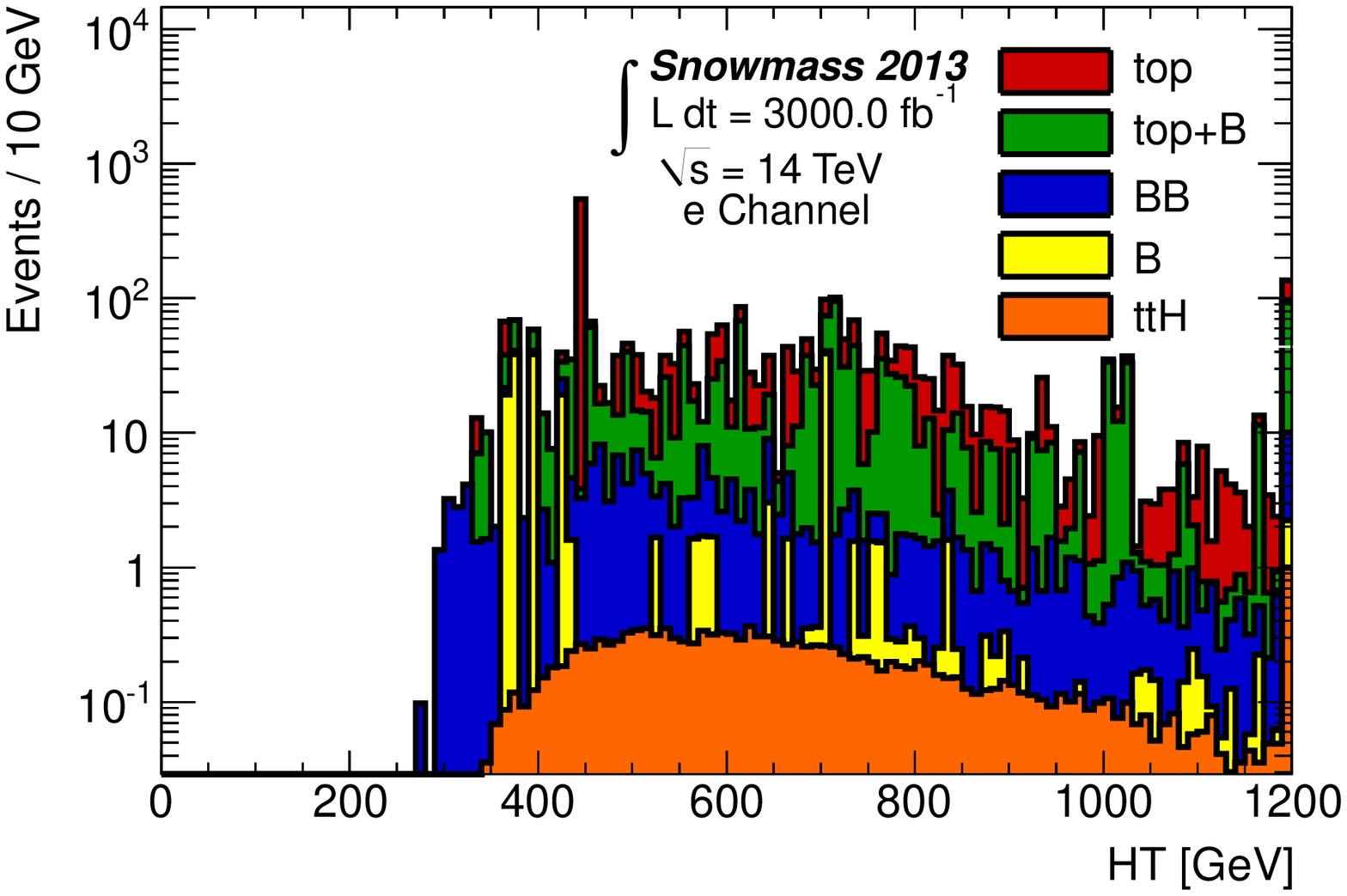}}}
%\caption{The \HT distribution for the muon (left) and electron (right) channels.}
\caption{Leading muon \pt and the \HT distributions for the muon and electron channels.}
\label{fig:precut}
\end{figure}

The \ttH, $\Hmumu$ candidate event is required to have exactly three
leptons (with at least 2 OS muons), at least one $b$-tagged jet, and
at least four jets in total. The leading lepton is typically a muon
from the Higgs boson decay, so to further reject background the
leading muon is required to have $\pt > 55$ GeV. This cut maintains
nearly full signal efficiency while cutting a respectable fraction of
background, as shown in Fig.~\ref{fig:precut}(a) and
Fig.~\ref{fig:precut}(b). Similarly, requiring the sum of lepton and
jet \pt (\HT) $>$ 350 GeV is also shown to cut background while
maintaining full signal efficiency in both channels and was thus
applied as shown in Fig.~\ref{fig:precut}(c) and Fig.~\ref{fig:precut}(d).
%%% Right now I say they are the same, I should probably show that they are. 

%Like the Higgs diphoton decay, we expect the Higgs dimuon decay to have excellent mass resolution.
In the electron channel, the two OS muons in the event are taken to
reconstruct the Higgs boson. For events in the muon channel, there are
two OS muon pairs to consider. We select the OS muon pair closest to
125 GeV as the Higgs boson candidate. A major background of \ttH where
Higgs boson decays to dimuons is \ttZ, which has a similar
signature. To suppress the $\ttbar Z$ background, which is
topologically similar to the signal, we reject an event if any OS
dimuon pair falls within the $Z$ window of 91-101 GeV.

The expected number of $\ttH$, $\Hmumu$ events in the full HL-LHC
dataset is about 400. When exactly 3 reconstructed leptons are
required in the event, the expected number of $\ttH$, $\Hmumu$ events
is reduced to about 54. The event selection is summarized in
Tables~\ref{tab:muchancuts}-~\ref{tab:elchancuts} along with cut
efficiencies for signal and background. The resulting signal
efficiency with respect to the preselection (`== 3 Leptons') is 57\% (
27\% and 30\% for the muon and electron channel, respectively).

\begin{table}[!htbp]
\centering
\begin{tabular}{ c | c | c | c | c} 
%\hline
Muon Channel & \ttH & \ttZ & \ttW & \tt \\
\hline
== 3 Leptons & --- & --- & --- & ---  \\
== 2 OS Muons + Muon & 48.8\% & 24.2\% & 9.36\% & 6.83\% \\
Leading Muon Pt $>$ 55 GeV & 97.8\% & 94.2\% & 79.0\% & 71.1\% \\
$\ge$ 1 tag & 78.5\% & 78.5\% & 86.7\% & 65.5\% \\
$\ge$ 4 jets & 83.9\% & 83.4\% & 76.9\% & 49.9\% \\
\HT $>$ 350 GeV & 99.1\% & 98.8\% & 100.\% & 91.7\% \\
No OS muon pairs in $Z$ window & 87.2\% & 17.6\% & 90.0\% & 53.4\% \\
\end{tabular}
\caption{Cut efficiencies for \ttH and selected major backgrounds. }
\label{tab:muchancuts}
\end{table}

\begin{table}[!htbp]
\centering
\begin{tabular}{ c | c | c | c | c} 
%\hline
Electron Channel & \ttH & \ttZ & \ttW & \tt \\
\hline
== 3 Leptons & --- & --- & --- & ---  \\
== 2 OS Muons + Electron & 49.8\% & 26.8\% & 32.0\% & 31.8\% \\
Leading Muon Pt $>$ 55 GeV & 95.2\% & 88.4\% & 73.9\% & 54.0\% \\
$\ge$ 1 tag & 78.1\% & 79.3\% & 83.3\% & 49.3\% \\
$\ge$ 4 jets & 83.4\% & 79.9\% & 80.0\% & 28.5\% \\
\HT $>$ 350 GeV & 99.3\% & 99.1\% & 100.\% & 99.6\% \\
No OS muon pairs in $Z$ window & 96.4\% & 17.5\% & 59.4\% & 51.8\% \\
\end{tabular}
\caption{Cut efficiencies for \ttH and selected major backgrounds. }
\label{tab:elchancuts}
\end{table}

%
%%%%%%%%%%%%%%%%%%%%%%%%%%%%%%%%%%%%%%%%%%%%%%%%%%%%%%%%%%%%%%%%%%%%%%%%%%%%%%%
% Analysis
%%%%%%%%%%%%%%%%%%%%%%%%%%%%%%%%%%%%%%%%%%%%%%%%%%%%%%%%%%%%%%%%%%%%%%%%%%%%%%%
%
\section{Analysis}

\begin{figure}[!htbp]	
\centering
\mbox{
\subfigure[\small Electron channel]{\includegraphics[scale=0.38]{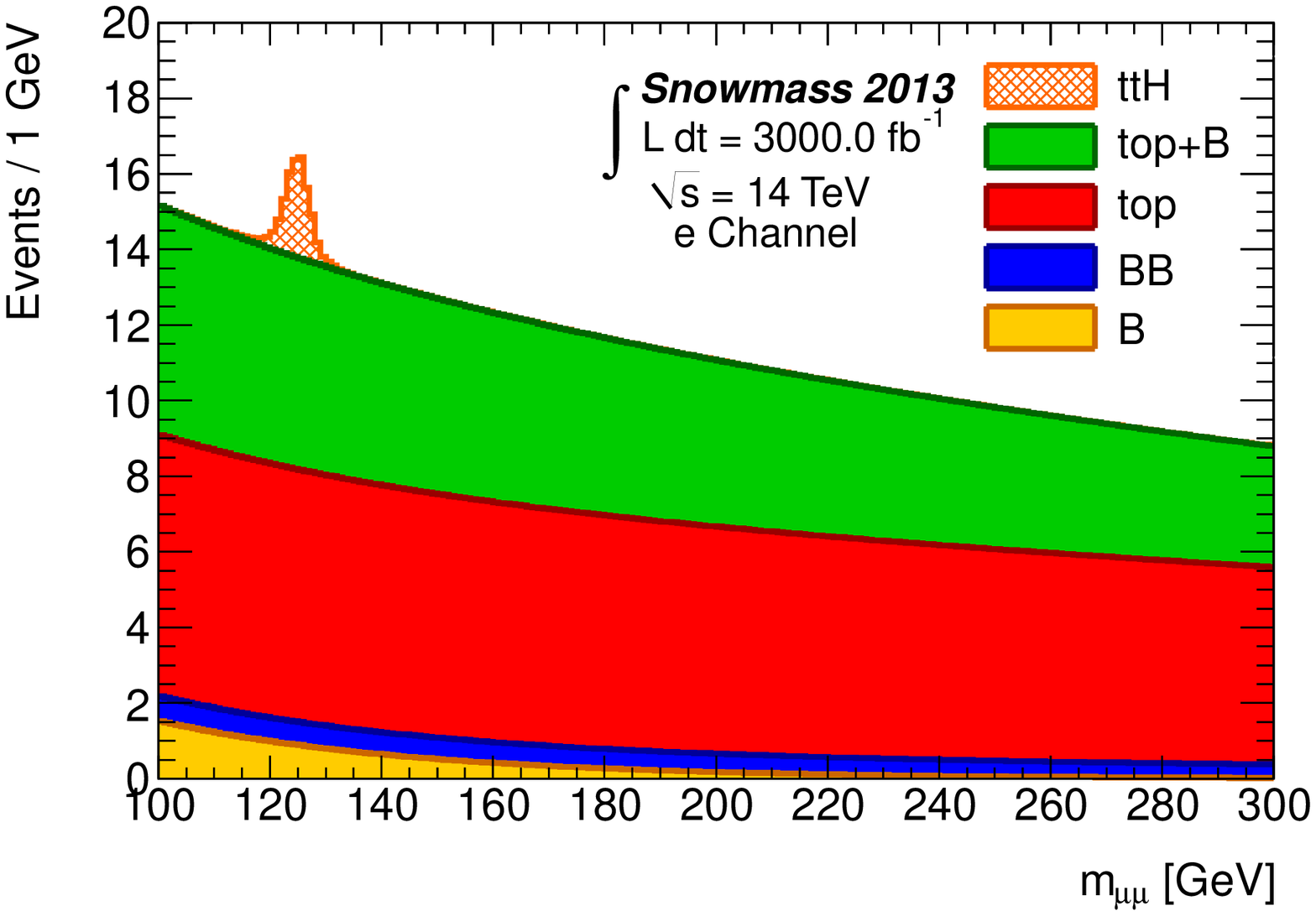}}
\subfigure[\small Muon channel]{\includegraphics[scale=0.38]{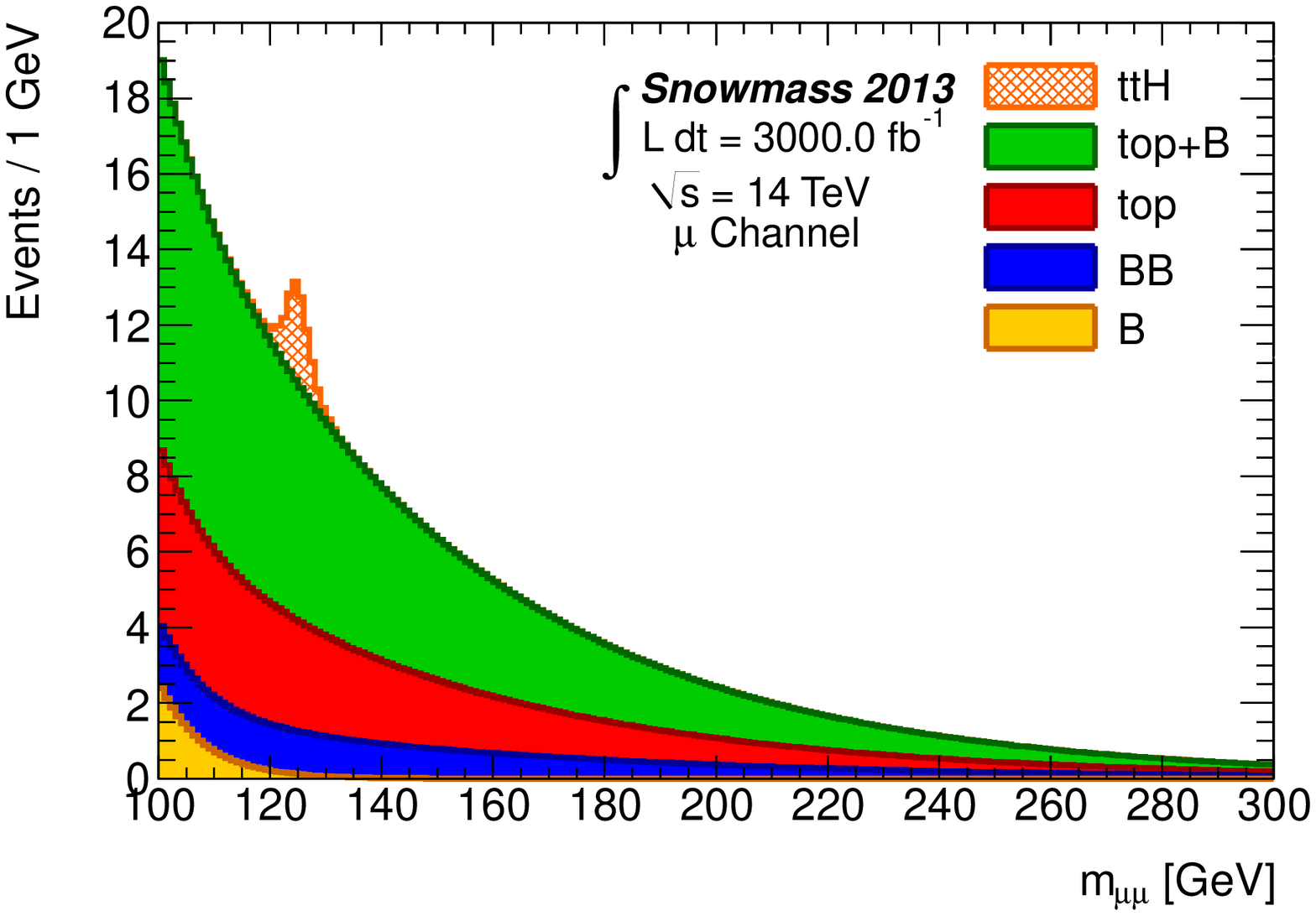}}}
\caption{Dimuon mass plot from the electron channel (a) and  muon channel (b) where signal is added to the background fit.}
\label{fig:mH_mny_e}
\end{figure}

\begin{figure}[!htbp]	
\centering
\includegraphics[scale=0.55]{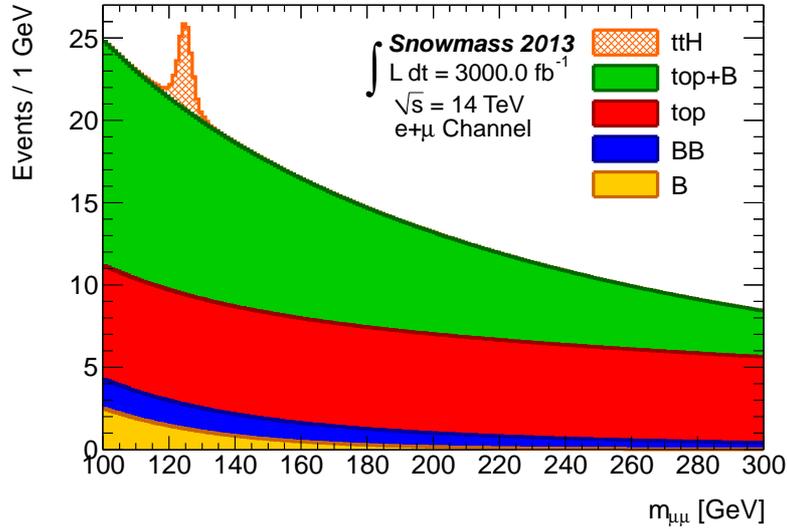}
\caption{Dimuon mass plot from the combined channel where signal is added to the background fit.}
\label{fig:mH_mny_c}
\end{figure}

To compensate for the statistical fluctuations in the background Monte
Carlo samples, we fit an exponential curve to each background using a
1 GeV binning and integrate over the mass window of 120-130 GeV. We
then count the signal events in the mass window to find
$S/\sqrt{B}$. Figures~\ref{fig:mH_mny_e}-~\ref{fig:mH_mny_c} show the
background fits and the expected SM Higgs boson signal.

%\begin{figure}[!htbp]	
%\centering
%\includegraphics[scale=0.55]{mH_mny_m}
%\caption{Dimuon mass plot from the muon channel where signal is added to the background fit.}
%\label{fig:mH_mny_m}
%\end{figure}

The expected numbers of events inside the Higgs boson mass window are shown in Table~\ref{tab:FinalNums}.

\begin{table}[!htbp]
\centering
\begin{tabular}{ c | c c c }
Channel &  Background [Events] & Signal [Events] & $S/\sqrt{B}$ \\
\hline
%\hline
Electron & 137.1 & 14.3 & 1.22 \\
Muon & 107.2 & 13.3 & 1.28 \\
\hline
Combined & 204.4 & 27.6 & 1.93 \\
\end{tabular}
\caption{Expected number of events for signal and background in each channel. }
\label{tab:FinalNums}
\end{table} 

%
%%%%%%%%%%%%%%%%%%%%%%%%%%%%%%%%%%%%%%%%%%%%%%%%%%%%%%%%%%%%%%%%%%%%%%%%%%%%%%%
% Conclusion
%%%%%%%%%%%%%%%%%%%%%%%%%%%%%%%%%%%%%%%%%%%%%%%%%%%%%%%%%%%%%%%%%%%%%%%%%%%%%%%
%
\section{Conclusion}

A precision measurement of the top Yukawa coupling is a key piece of
any further understanding of electroweak symmetry breaking and the
origin of mass. Due to the clean event topology, the \ttH ($\Hmumu$)
production and decay would be an excellent signature in which to make
such studies. We find that the low branching ratio for Higgs boson
decay to muons, combined with the small \ttH production cross section,
make such analysis in the trilepton signature extremely difficult,
even with the ultimate expected luminosity (3000 fb$^{-1}$) at the
HL-LHC.

Considering only statistical uncertainties, the expected sensitivity
to the Standard Model \ttH, $\Hmumu$ production and decay is at the
level of $2\sigma$. To claim the evidence of the $\ttH$, $\Hmumu$
signal one would need to use multivariate techniques and/or study the
signature in the all-hadronic $\ttbar$ decays.

%%%%%%%%%%%%%%%%%%%%%%%%%%%%%%%%%%%%%%%%%%%%%%%%%%%%%%%%%%%%%%%%%%%%%%%%%%%%%%%
% Appendices
%%%%%%%%%%%%%%%%%%%%%%%%%%%%%%%%%%%%%%%%%%%%%%%%%%%%%%%%%%%%%%%%%%%%%%%%%%%%%%%

%\appendix
%\part*{Appendices}
%\addcontentsline{toc}{part}{Appendices}
%\input{appendix_mc.tex}

%%%%%%%%%%%%%%%%%%%%%%%%%%%%%%%%%%%%%%%%%%%%%%%%%%%%%%%%%%%%%%%%%%%%%%%%%%%%%%%
% Bibliography
%%%%%%%%%%%%%%%%%%%%%%%%%%%%%%%%%%%%%%%%%%%%%%%%%%%%%%%%%%%%%%%%%%%%%%%%%%%%%%
%
% Style file to use with mcite.
% Use atlasstyle with just cite.
%\bibliographystyle{atlasstylem}

%\clearpage

%\bibliographystyle{atlasstyle}
\small
\bibliography{ttH_mumu}{}
\end{document}